\documentclass[11pt,a4paper]{article}
\usepackage{jheppub}

\usepackage{epsfig}
\usepackage{latexsym}
\usepackage{amsfonts}
\usepackage{amsmath}
\usepackage{amsthm}
\usepackage{amssymb}
\usepackage{amsbsy}
\usepackage{multirow}
\usepackage{slashed}
\usepackage{amsmath,latexsym,amssymb}
\usepackage{graphicx}
\usepackage{psfrag}
\usepackage[latin1]{inputenc}
\usepackage{nicefrac}
\usepackage[stable]{footmisc}
\usepackage{braket}
\usepackage{rotating}
\usepackage{afterpage}
\usepackage{bbm}

\def\be{\begin{equation}}
\def\ee{\end{equation}}
\def\bea{\begin{eqnarray}}
\def\eea{\end{eqnarray}}
\def\ba{\begin{eqnarray}}
\def\ea{\end{eqnarray}}

\newcommand{\stln}{\setlength{\unitlength}{2.2ex}}

\newcommand{\sfr}{\framebox(1,1){\begin{picture}(1,1)
  \put(0,0){\line(1,1){1}}\end{picture}}}

\newcommand{\sgenrowbox}
{\stln \lower1.4ex\hbox{
\begin{picture}(7.6,1.6)
\multiput(.3,.3)(1,0){3}{\sfr}
\put(3.3,.3){\framebox(3,1){$\cdots$}}
\put(6.3,.3){\sfr}
\end{picture}}}

\newcommand{\ft}[2]{{\textstyle\frac{#1}{#2}}}

\newcommand{\cn}{{\cal N}}

\def\be{\begin{equation}}
\def\ee{\end{equation}}
\def\bea{\begin{eqnarray}}
\def\eea{\end{eqnarray}}
\def\ba{\begin{array}}
\def\ea{\end{array}}
\def\bd{\begin{displaymath}}
\def\ed{\end{displaymath}}

\def\a{\alpha}
\def\b{\beta}

\def\g{\gamma}

\def\>{\rangle} 
\def\<{\langle} 
\def\Dsl{D \hskip-.6em \raise1pt\hbox{$ / $ } }

\newcommand{\eps}{\epsilon}

\newcommand{\eq}{\begin{equation}}
\newcommand{\en}{\end{equation}}

\newcommand{\beq}{\begin{equation}}
\newcommand{\eeq}{\end{equation}}

\newsavebox{\uuunit}
\sbox{\uuunit}
    {\setlength{\unitlength}{0.825em}
     \begin{picture}(0.6,0.7)
        \thinlines
        \put(0,0){\line(1,0){0.5}}
        \put(0.15,0){\line(0,1){0.7}}
        \put(0.35,0){\line(0,1){0.8}}
       \multiput(0.3,0.8)(-0.04,-0.02){12}{\rule{0.5pt}{0.5pt}}
     \end {picture}}

\def\be{\begin{equation}}
\def\ee{\end{equation}}
\def\bea{\begin{eqnarray}}
\def\eea{\end{eqnarray}}

\def\a{\alpha}
\def\b{\beta}
\def\g{\gamma}

\def\E {$E_{7(7)}$}

\def\sF{{{ F}\!\!\!\!\hskip.8pt\hbox{\raise1pt\hbox{/}}\,}}
\def\som{{{ \omega}\!\!\!\!\hskip.8pt\hbox{\raise1pt\hbox{/}}\,}}
\def\sJ{{{\rm J}\!\!\!\!\hskip.8pt\hbox{\raise1pt\hbox{/}}\,}}

\def\a{\alpha}
\def\b{\beta}
\def\ba{\bar \alpha}

\def\eps{\epsilon}
\def\g{\gamma}

\newcommand{\cN}{\mathcal{N}}
\newcommand{\cG}{\mathcal{G}}
\newcommand{\cH}{\mathcal{H}}
\newcommand{\rf}[1]{(\ref{#1})}

\title{{ { Supersymmetry constraints on U-duality invariant deformations   of {\boldmath$\cN \geq 5$} Supergravity} }}
  \author[a,b]{Murat  Gunaydin,}    \author[a]{Renata  Kallosh}

\affiliation[a]{Stanford Institute for Theoretical Physics and Department of Physics, Stanford University,\\ Stanford, CA 94305-4060, USA}
\affiliation[b]{Institute for Gravitation and the Cosmos and Department of Physics, Pennsylvania State University, \\ University Park, PA 16802, USA
}
\emailAdd{mgunaydin@psu.edu}  \emailAdd{kallosh@stanford.edu}

\abstract{Candidate counterterms break
 E7   type U-duality symmetry of $\cN \geq 5$ supergravity theories  in four dimensions \cite{Kallosh:2011dp}.  A proposal was made in \cite{Bossard:2011ij} to restore it, starting with a double set of vector fields and argued that a supersymmetric extension of their proposal should exist.
 We show that the extra vectors, needed for the  deformation,  can not be auxiliary fields in an eventual  off-shell formulation $\cN \geq 5$ supergravity, assuming that such a formulation exists.
 Furthermore we show that these extra vector fields can not be dynamical either since that changes  the unitary supermultiplets underlying these theories and requires one to go  beyond the standard framework of extended simple supergravities. To show this
we list all relevant unitary conformal supermultiplets of $SU(2,2|\cN+n)$. We find that doubling  of vectors consistent with linearized supersymmetry  requires to change the number of  scalars, violating the coset structure of the theory,  and also to add a finite number of  higher spin fields,  which do not  admit consistent couplings to theories with spins $\leq 2$.
Thus, the proposed duality  restoring deformation along the lines of \cite{Bossard:2011ij} can not be implemented within the standard framework of extended supergravity theories.
We argue therefore that, in the absence of anomalies,  E7 type duality together with supersymmetry,  might protect $\cN \geq 5$ supergravity from UV divergences, in particular,   $\cN=5$ supergravity at 4 loops in d=4.}

\begin{document}
 \maketitle
\section{Introduction}
The  3-loop and 4-loop UV finiteness \cite{Bern:2007hh}  of $\cN=8$ supergravity in $4d$
\cite{Cremmer:1979up}
was explained in various ways, either using the absence of candidate counterterms in the light-cone formalism \cite{Kallosh:2008mq} or in a Lorentz covariant case, based  on \E\, duality symmetry and supersymmetry in  \cite{Brodel:2009hu,Beisert:2010jx} and  in \cite{Kallosh:2011dp,Bossard:2011tq}.
 These explanations were  based on the expected structure of the Lorentz covariant candidate counterterms  \cite{Kallosh:1980fi} and supergravity duality symmetry \cite{Cremmer:1979up,Gaillard:1981rj}.

 Most relevant to the current work is the observation  that all known counterterms, candidates for UV divergences in N = 8 supergravity,  would break the Noether-Gaillard-Zumino \E\,  deformed duality current conservation \cite {Kallosh:2011dp}. However, Bossard and Nicolai (BN) suggested  in \cite{Bossard:2011ij}  that it is possible to fix the problem pointed out in \cite {Kallosh:2011dp} by deforming  the classical  twisted self-duality constraint in the presence of higher derivative terms in the action.

 In the classical maximal supergravity theory  with \E\ symmetry  there are 28 independent Maxwell field strengths as a consequence of supersymmetric twisted self-duality constraint \cite{Cremmer:1979up}.
To identify a deformed constraint according to BN, one has to find a manifestly duality invariant higher derivative supersymmetric invariant which requires doubling of the Maxwell field strengths and their duals, i. e.  two sets transforming in $\mathbf{28}$ and $\overline{\mathbf{28}}$ of the R-symmetry group $SU(8)$
 in $\cN=8$ supergravity. Examples of such deformation of the
classical  twisted self-duality constraint were given in  \cite{Bossard:2011ij} for some non-supersymmetric models. The proposal in  \cite{Bossard:2011ij} was further developed  in \cite{Carrasco:2011jv} where  covariant procedures for perturbative non-linear deformations of duality-invariant theories were established to all orders in the deformation parameter. The
starting point requires the existence of some `Source of Deformation' (SoD).   Various examples of SoD's were given in \cite{Carrasco:2011jv} which resulted in building novel models with $U(1)$ duality symmetry.

It was actually known long before the BN supergravity proposal \cite{Bossard:2011ij},  that perturbative non-linear deformations of $U(1)$ duality-invariant theories with  global
 supersymmetry are available, see for example \cite{Kuzenko:2000uh} where the review of
nonlinear selfduality and supersymmetry is given, and many aspects of related problems were studied much earlier.

More recently a deformation proposal was presented  in \cite{Kallosh:2018mlw} for $\cN\geq 5$ supergravity, using a symplectic formalism of \cite{Andrianopoli:1996ve} that was developed for studies of black hole attractors. A closed form of the bosonic  deformed action, exact  to all orders in deformation parameter was produced for an original choice of the SoD in \cite{Bossard:2011ij}. The issue of a supersymmetrization of such a SoD remains unclear.

Using the first order formalism with manifest E7 duality symmetry an explanation of  the UV finiteness at L=4, $\cN=5$, discovered in \cite{Bern:2014sna},  was proposed in \cite{Kallosh:2018wzz}.   Assuming absence of duality-supersymmetry anomalies, it was argued in \cite{Kallosh:2018wzz} that these symmetries might protect $\cN\geq 5$ at higher loops\footnote{ It was shown in \cite{Bossard:2010dq} that $E7$ duality symmetry can be maintained at all orders of perturbation theory using a non-Lorentz covariant formulation of the theory. See the discussion section on this point.}. Here we will reach the same conclusion based on supersymmetry in the framework of the second order formalism, in particular with regard to UV finiteness at L=4, $\cN=5$.

It was assumed in BN proposal in \cite{Bossard:2011ij} that the supersymmetric version of the proposal for SoD is available,  and that all other symmetries of the theory, local supersymmetry and  general covariance  are respected.
The existence of the supersymmetric SoD in $\cN=8$ supergravity was further investigated in  \cite{Kallosh:2012yy} and it  was concluded   there  that its existence  with ${\bf 56}$ independent vectors contradicts the $\cN=8 $ superspace construction \cite{Brink:1979nt},  and the relevant solutions of the superspace Bianchi identities. It was suggested in \cite{Kallosh:2012yy} that the existing $\cN=8$ superspace has to be  deformed to admit the SoD. The superspace deformation is
very complicated and things remained inconclusive.

The supersymmetry issue was investigated in \cite{Gunaydin:2013pma}  at the linearized level, using the fact that  the linearized $\cN=8$ supergravity is based on the representations of $SU(2,2|8)$ superconformal algebra. More specifically it was shown long ago that the fields of $\cN=8$ supergravity can be fitted into the CPT self-conjugate doubleton supermultiplet  of the $\cN=8$ superconformal algebra $SU(2,2|8)$\cite{Gunaydin:1984vz}. Motivated by the work of \cite{Beisert:2010jx} on  the study of counterterms of maximal supergravity using this doubleton supermultiplet it was reformulated in terms of constrained superfields in \cite{Chiodaroli:2011pp}.  The result of the investigation of  \cite{Gunaydin:2013pma} was that
 it is not possible to deform the maximal supergravity to restore \E\ duality, while maintaining both general covariance and $\cN=8$ supersymmetry, as was proposed in  \cite{Bossard:2011ij}, if the required extra vector fields  are assumed to be dynamical. Deformation of  $\cN=8$ supergravity  with extra dynamical vector fields along the lines of BN proposal necessarily involves higher spins and multiple gravitons and hence requires one to go beyond the standard framework of supergravity.

 More recently, a significant progress was achieved in understanding the absence of anomalies in one-loop amplitudes in $\cN\geq 5$ supergravities  \cite{Freedman:2017zgq}. The new analysis in  \cite{Freedman:2017zgq} is based on a classification of all linearized chiral superfields in $\cN\geq 5$ supergravities, in addition to constrained ones which were known for a long time  \cite{Kallosh:1980fi}. It became possible to explain the UV finiteness of $L=3=\cN-2$ in  $\cN=5$  theory \cite{Freedman:2018mrv}, using the soft scalar limits of amplitudes, as it was done for  $L\leq 6=\cN-2$ in  $\cN=8$ case in  \cite{Beisert:2010jx}, but not in the case of $L=4=\cN-1$, $\cN=5$  theory. This is analogous to the case $L= 7=\cN-1$  in $\cN=8$ theory,   where the soft scalar limit analysis was not conclusive.

Here we will first revisit the  work of  \cite{Gunaydin:2013pma} and extend and apply the  analysis given there  to all supergravities with $\cN \geq 5$. The BN proposal required that 1) the number of vectors is doubled to make $\cG$ duality manifest, but 2)  scalars have to preserve the original ${\cG\over \cH}$ coset space. As was shown for the $\cN=8$ supergravity in \cite{Gunaydin:2013pma} we find that the doubling of vectors requires the introduction of higher spins and multiple gravitons for all $\cN \geq 5$ supergravities if the extra vector fields are dynamical. Furthermore we show that this doubling process with extra dynamical vectors necessarily introduces additional scalar fields which changes the coset ${\cG\over \cH}$. This means that  the BN proposal with extra dynamical vector fields can  not be made supersymmetric within the standard framework of  supergravity with a single graviton.

This raises the question whether  the BN proposal can be  consistent with supersymmetry if the extra vector fields are not dynamical. In particular,  one may ask whether one can use auxiliary vector fields in an off-shell  formulation of $\cN \geq 5$ supergravities to drive the deformation. To date no off-shell formulations of $\cN \geq 5$ supergravities have been found. However $\cN \geq 5$ supergravities admit consistent truncations to $\cN=2$ supergravity coupled to vector multiplets whose off-shell formulations are known. Using this fact  we will also show that the extra vector fields needed for the deformation can not be identified with some auxiliary fields in a possible off-shell formulation of $\cn \geq 5$ supergravities.

The plan of the paper is the following.
In Sec. 2 we review the BN proposal about the deformation of the supergravity action based on a SoD. In Sec. 3,
 we study the question whether the extra vector fields needed for the deformation can be identifed  with auxiliary fields in  potential off-shell formulation of $\cN \geq 5$ supergravity theories and argue these extra vector fields   can not be identified with some of the auxiliary fields.
In Sec. 4 we explain how using the unitary representations of the conformal superalgebras $SU(2,2|\cN +n)$ we can list all possible supermultiplets which have a chance to support the BN proposal with extra  dynamical vectors.
We therefore will list all possibilities to double the number of physical vectors such that  the linearized approximation preserves the $SU(2,2|\cN)$ superconformal symmetry. For $\cN=8$ it was done in  \cite{Gunaydin:2013pma} but here we will pay attention to the number of scalars in supersymmetric theories with a double number of vectors and extend the analysis to the case of $\cN=6,5$. In all cases we will have two options. In the first case, in Sec. 5  we do not  enlarge the R-symmetry group and try to find all possible supermultiplets which will allow us to double the vectors within $SU(2,2|\cN)$ for the $\cN$-extended supergravity. In the second case in Sec. 6 we enlarge the R-symmetry group, corresponding to increase in the number of $Q, S$ supersymmetries,  and study multiplets of $SU(2,2|\cN+n)$ superalgebra with a consequent decomposition into $SU(\cN)\times SU(n)$. We also present the corresponding supermultiplets as linearized superfields, in Sec. 7. In  the Appendix we discuss the related issues for $\cN=2$ supergravity with matter.

We conclude that in all cases, when doubling vectors, we either have to double the scalars in their  required $SU(\cN)$ representation, or we have to add  $SU(\cN)$ singlet scalars.  These changes in the number of scalar fields are incompatible with duality symmetry since the coset space structure is not preserved. Also,  we always get some fields in these supermultiplets  with higher spins $s>2$,  which makes the coupling to gravity questionable \cite{Gunaydin:2013pma}. Furthermore they contain multiple gravitons whose interacting theories have been shown to be inconsistent \cite{Boulanger:2000rq}.
Therefore the proposal of \cite{Bossard:2011ij} to restore duality symmetry, broken by UV divergence, can not be made consistent with supersymmetry within the standard supergravity framework. In the absence of any other  proposal to do it, we conclude that the UV finiteness in $\cN=5$ supergravity at 4 loops may be explained by the fact that UV divergence breaks duality symmetry \cite{Kallosh:2011dp}.  Whether  a BN type proposal  \cite{Bossard:2011ij}  to restore duality symmetry in presence of  UV divergences can be made consistent with supersymmetry  beyond the standard framework of supergravity, such as string theory or higher spin theories is an open problem.

\section{Source of Deformation Proposal}

Here we explain why  a doubling of vectors is required in the SoD according to BN proposal in $\cN=8$, and we generalize it to the case of $\cN\geq 5$. So we will need $\mathbf{56,32,20}$ instead of $\mathbf{28,16,10}$ physical vectors for $\cN=8,6,5$ respectively. We will also explain here that SoD according to BN proposal  has to depend on original scalars forming a coset space ${\cG\over \cH}$. So, we need to confirm that the numbers of scalars remains equal to the number of physical scalars, coordinates of
a coset space ${\cG\over \cH}$, and moreover, they must transform in the representations of the $\cH$ group as to represent the coordinates of the coset space ${\cG\over \cH}$. It means, we need to recover after doubling of vectors, the required number of  scalars is: $\mathbf{70}$ in $SU(8)$, $\mathbf{15}$ and $\overline {\mathbf{15}}$ in $SU(6)$ and $\mathbf{5}$ and $\overline {\mathbf{5}}$ in $SU(5)$ for $\cN=8,6,5$ respectively.

Using notation of \cite{Kallosh:2018mlw, Andrianopoli:1996ve}  we introduce  a $2 n_{v}$-dimensional real symplectic
vector of field strengths $\mathcal{F}$ and a symplectic section $\mathcal{V}_{AB}$ describing the scalars of the theory

\begin{equation}
\mathcal{F} \equiv
\left(
  \begin{array}{c}
  F^{\Lambda} \\ G_{\Lambda}  \\
  \end{array}
\right)\,  , \qquad \mathcal{V}_{AB}
\equiv
\left(
  \begin{array}{c}
  f^{\Lambda}{}_{AB} \\ h_{\Lambda\, AB} \\
  \end{array}
\right)= - \mathcal{V}_{BA}\, .
\label{doubletF}\end{equation}
$\mathcal{F}$ transforms in the $\mathbf{56,32,20}$ of the corresponding duality groups of type E7, namely,  \E\,, $SO^*(12)$  and  $SU(1,5)$ for $\cN=8,6,5$ respectively \cite{Br}.
The scalars of the theory are coordinates of the ${\cG\over \cH}$ coset space where $\cG$ is type E7 group and $\cH$ is an isotropy group, $SU(8), SU(6)\times U(1), SU(5)\times U(1)$ for $\cN=8,6,5$ respectively and
where the pair of indices $A,B=1,\cdots,\cN$  are raised
and lowered by complex conjugation. The period matrix is
$
h_{\Lambda\, AB} = \mathcal{N}_{\Lambda\Sigma}f^{\Sigma}{}_{AB}
$.
The graviphoton field strengths are given by the symplectic invariant,
\begin{equation}
\label{eq:graviphotondef}
T_{AB}
\equiv
\langle \mathcal{V}_{AB}\mid \mathcal{F} \rangle \equiv  \mathcal{F}^\Lambda   \mathcal{V}_{\Lambda AB}- \mathcal{F}_\Lambda   \mathcal{V}^{\Lambda}_{ AB} =- T_{BA}
\, .
\end{equation}
Graviphotons are invariants of the global U-duality group and  transform covariantly under compensating $SU(\cN)$ transformations. In classical $\cN\geq 5$ supergravity, in the absence of fermions, there is a  linear twisted self-duality constraint:
\begin{equation}
\label{linear}
{T}_{AB}{}^{ +} =  h_{\Lambda
 AB}\, F^{+\Lambda}_{\mu\nu} - f^\Lambda_{AB} \,G_{ \mu\nu\,\Lambda}^+ =0 \, .
\end{equation}
Here
 self- and anti-selfdual parts of $T$ are
$
T_{AB}{}^{\pm}
\equiv
\langle \mathcal{V}_{AB}\mid \mathcal{F}{}^{\pm}   \rangle$,  $  {T}^{* \pm AB}
\equiv
\langle \overline {\mathcal{V}}^{AB}\mid \mathcal{F}{}^{\pm}   \rangle
$.
The constraint  \rf{linear} results in the relation between $G$ and $F$, so that only one of them is independent
\be
G^+= \cN F^+\ ,  \qquad G^-= \overline \cN F^- \ .
\label{GofF}\ee
This gives a correct amount of the physical degrees of freedom for vector fields , $\mathbf{28,16,10}$  and is  one-half of the symplectic representation of the E7-type symmetry  for \E\, , $SO^*(12)$ and $SU(1,5)$ duality, respectively.

When the candidate UV divergences are added to the classical action, bosonic  linear twisted self-duality constraint \rf{linear} is deformed, following \cite{Bossard:2011ij}.  The new constraint  can be given in an  $\cH$-covariant form as proposed in \cite{Carrasco:2011jv}
\begin{equation}
\label{con}
{T}_{AB}{}^{ + \, def }\equiv  {T}_{AB}{}^{ +} - \lambda {\delta I (T^-,  T^{*+})\over \delta T^{*+AB}}=0 \ .
\end{equation}
 Here the source of deformation
\be
I (T^-,  T^{*+})
\label{source}\ee
depends on a doublet of vector fields $\mathcal{F}$ shown in eq. \rf{doubletF} where there is no relation between the upper and  lower components of a doublet, $F$  and $G$  as given for example in the classical case in eq. \rf{GofF}. $\cH$ covariance of the constraint in \rf{con} means covariance in  an isotropy group, $SU(8), U(6), U(5)$ for $\cN=8,6,5$ respectively.

\

The doubling of independent vectors versus physical ones  is a cornerstone in the \E\, duality covariant form of the deformation in $\cN=8$ theory proposed in \cite{Bossard:2011ij}. In their notation a duality doublet $F^m$ consists of two sets of 28 Maxwell field strengths
\be
F^m \equiv (F^{\textsl{a}}, F^{\bar {\textsl{a}} })\, , \qquad \textsl{a}=1,...,28 \qquad  \bar {\textsl{a}} =1,...,28
\label{doublet}\ee
In case of $\cN=8$ this equation is manifestly \E\, invariant if the SoD is a duality invariant functional depending on a duality doublet  (\ref{doublet}) where the two sets of 28 vectors are  independent. The vector part of SoD is
\be
{\cal I}(F^m) = {\cal I} [ F^{\textsl{a}}, F^{\bar {\textsl{a}} }] \ .
\label{SoD}\ee
The proposal for the vector part of the deformed twisted self-duality constraint in notation of \cite{Bossard:2011ij} is given by
\be
F^m + J^m{}_n \tilde F^n= G^{mn} {\delta{\cal I} \over \delta F^n} + \Omega^{mn} {\delta{\cal I} \over  \delta \tilde F^n}\,  ,  \qquad m=1,...,56
\label{BN}\ee
Here for $\cN=8$ supergravity $G_{mn}$ is a scalar dependent symmetric metric $G_{mn} \in  E_{7(7)} \subset Sp(56, \mathbb{R})$. $J^m{}_n$ is a `complex structure' ,  $\Omega^{mn}$ is a symplectic form and $\tilde F_{\mu\nu} = {1\over 2\sqrt{-g} }\epsilon_{\mu\nu}{}^{\rho\sigma} F_{\rho\sigma}$ is dual to $F$. The classical twisted self-duality constraint, in the absence of deformation, is
$
F^m + J^m{}_n \tilde F^n=0
$,
which is a relation expressing one of the 28 via the other, so that the theory has only one set of 28 vectors in agreement with unitarity. This is an analog of eq. \rf{linear} where the constraint is $SU(8)$ covariant, \E\, invariant. To deform it according to (\ref{BN}) the SoD action ${\cal I}[ F^{\textsl{a}}, F^{\bar {\textsl{a}} }] $ has to be differentiated over the set of 56 independent vectors.

The proposal in \cite{Bossard:2011ij} requires that the number of vectors is doubled but the number of scalars remains the same, in general,  since the coset space ${\cG\over \cH}$ is the same. However, in practice when the proof of consistency of the proposal is given order by order in deformation in Appendix of \cite{Bossard:2011ij}, it is given only for vanishing scalars.

\section { Off-shell $\cN\geq 5$ supersymmetry?}

It is generally  believed that any off-shell formulation of $\cN\geq 5$ supergravity, if it exists, requires an infinite number of auxiliary fields.
Assuming that an off-shell formulation of $\cN\geq 5$ supergravity exists one may then pose  the  question whether  the second set of vector fields required by the BN proposal, which form a symplectic doublet, could be identified with some of the auxiliary vector fields. If that were the case, when counting physical degrees of freedom in on shell multiplets,  we would not have to double the set of dynamical vector fields in agreement with supersymmetry. We will now present arguments against such a possibility, based on all known supersymmetry constructions.

 $\cN =8$ supergravity can be truncated consistently to an $\cN=2$ Maxwell-Einstein supergravity theory describing the coupling of 15 vector multiplets which is known as the quaternionic magical supergravity theory \cite{Gunaydin:1983rk,Gunaydin:1983bi}. The quaternionic magical supergravity has the same bosonic field content as the $\cN=6$ supergravity but with a different fermionic spectrum. It has $SO^*(12)$ as its U-duality group under which 16 vector field strengths, including the graviphoton, and their magnetic duals transform in the 32 dimensional spinor representation.
 Therefore the BN analysis applied to the bosonic sector of $\cN=6$ supergravity would yield  identical results as  the BN scheme applied to the purely bosonic sector of $\cN=2$ quaternionic magical  supergravity theory. We do not know the auxiliary fields of $\cN=6$ supergravity. However the off-shell formulation of $\cN=2$ supergravity and its couplings to vector multiplets are known.

 If a supersymmetric deformation of maximal supergravity existed sourced by the auxiliary fields in an off-shell formulation , that preserves E7 duality, one can truncate it to the deformed quaternionic magical supergravity theory that preserves both $\cN=2$ supersymmetry and $SO^*(12)$ duality symmetry. However off-shell formulation of $N=2$ quaternionic magical supergravity can not have auxiliary vector fields transforming in the $15+1$ dimensional representation of the isotropy group $U(6)$ of its scalar manifold $SO^*(12)/U(6)$. This follows from the fact that the Weyl multiplet of $\cN=2$ supersymmetry has auxiliary vector fields transforming in the adjoint representation $3+1$ of the R-symmetry group $U(2)$  and the $\cN=2$  vector supermultiplets do not have any auxiliary vector fields and contain only auxiliary scalar fields \footnote{Off-shell formulation of $\cN=2$ supergravity has a long history. We refer the reader to relatively recent papers \cite{Banerjee:2011ts,deWit:2010za} and the references therein.}.
 The only other massless $\cN=2$ supermultiplet that contains vector fields is the $s=3/2$ gravitino multiplet. It has recently been shown that the $s=3/2$ Rarita-Schwinger field can be coupled to vector fields  consistently only as part of a graviton supermultiplet and that the $\cN=2$ pure supergravity is unique \cite{Boulanger:2018fei}.  These results show clearly that it is impossible to deform the maximal supergravity consistent with supersymmetry sourced by auxiliary vector fields along the lines of BN proposal.

 In \cite{Chemissany:2012pf} the $\cN=2$ supergravity was studied in the off-shell superconformal framework with all auxiliary fields present. Consistent with earlier results in \cite{Banerjee:2011ts,deWit:2010za} the only auxiliary vector  fields appear in the adjoint of the  R-symmetry group $U(2)$.
Furthermore the authors of   \cite{Chemissany:2012pf} show that the supersymmetry rules of a classical supersymmetric off-shell theory are inconsistent with the UV divergences and have to be deformed. Such a deformation of a superspace was never performed and it is not known if a consistent version of it is even  possible. Therefore the analysis in \cite{Chemissany:2012pf} raises serious issues, independent of the BN proposal, about the compatibility of candidate counterterms in higher $\cN$ supergravities with off-shell supersymmetry at the non-linear level.

We stress here that the possibility to build the symplectic multiplet from one half physical vectors and one half auxiliary vectors vanishing on shell, is not attractive since duality symmetry mixes them,  and equations of motion get mixed with Bianchi identities.
In fact  an independent argument as to why the extra field strength $G_{\mu\nu}$  in \rf{doubletF}, independent of $F_{\mu\nu}$,
required in BN SoD,  can not be auxiliary, not propagating,  can be given as follows. Once the (bosonic) deformation of $\cN\geq 5$ supergravity is achieved, for example in the case studied in  \cite{Kallosh:2018mlw} and the relation between the $F$ and $G$ component of the symplectic doublet in \rf{doubletF} is established, one finds that
\be
G^+ = -i F^+ + \cdots
\ee
Here terms with $\cdots$ include higher order terms in gravitational coupling $\kappa$ and in parameter of deformation $\lambda$,
as one can see from  eq. (3.9) in  \cite{Kallosh:2018mlw}. These terms are non-linear in fields. Thus in approximation that $\kappa=\lambda=0$, $G_{\mu\nu}$ is proportional to $F_{\mu\nu}$ and if $G_{\mu\nu}$  would be an auxiliary field it would be impossible for it also to be proportional to a physical field, up to non-linear terms. This argument, by itself, appears to be sufficient to rule out an attempt to use auxiliary field for the $G$ part of the doublet. In any case, here we have shown technically, using a consistent truncation to  $\cN=2$ Maxwell-Einstein supergravity, that there are no auxiliary vector fields with required transformation properties in these theories which can be used for the BN type deformation.

Therefore we will study below all options to realize linear supersymmetry with the doubled number of dynamical vectors in the multiplets.

\section{Source of Deformation and Supersymmetry}

The BN proposal for maximal supergravity and its extensions to $\cN=6$ and $\cN=5$ supergravity with a source depending on vectors and scalars, as shown in eq. \rf{source} was assumed to have a supersymmetric extension with $\cN$ supersymmetries. The candidate counterterms for these supergravity theories were constructed starting with  \cite{Kallosh:1980fi} using linearized superfields corresponding to physical states of the theory. In particular, such superfields depend on vector field strengths in $\mathbf{28,16,10}$  and $\mathbf{\overline {28},\overline {16},\overline {10}}$ in $SU(8), SU(6), SU(5)$   representations in $\cN=8,6,5$ respectively.

In general, the classification of the
  massless unitary supermultiplets of extended Poincar\'e superalgebras is well known  \cite{Nahm:1977tg},  \cite{Strathdee:1986jr}.
   The minimum spin range for the massless unitary supermultiplets of $\cN$-extended Poincar\'e superalgebras is $\frac{\cN}{4}$ for even $\cN$ and the maximum number of Poincare supersymmetry generators is 32.   It was shown in \cite{Gunaydin:1984vz} that  the fields of maximal supergravity in $d=4$ can be fitted into an ultra short CPT-self-conjugate unitary supermultiplet ( doubleton)  of the conformal superalgebra $SU(2,2|8)$ with 64 supercharges. Even though the fields of $\cN=8$ supergravity form a representation of the $\cN=8$ conformal superalgebra interactions of the maximal supergravity break the conformal supersymmetry down to its Poincare subsuperalgebra. The corresponding supermultiplet can be written as a linear constrained superfield \cite{Chiodaroli:2011pp}. These superfields have been used in the analysis of counterterms in maximal supergravity \cite{Brodel:2009hu,Beisert:2010jx}.   Since the superfields used in writing down linearized counterterms in $\cN\geq 5$ supergravity correspond to conformal supermultiplets we will perform our analysis of conformal superalgebras $SU(2,2|\cN)$ in four dimensions using the oscillator method \cite{Gunaydin:1984fk,Gunaydin:1984vz,Gunaydin:1998sw,Gunaydin:1998jc}. However, as different from the standard superfields that enter in the counterterms, we will be looking for different  supermultiplets that can couple to and extend the maximal supergravity with the total  number of vector fields doubled. We will also check if it is possible to find the suitable supermultiplets by  embedding them  into larger $SU(2,2|\cN +n)$ conformal superalgebras with the consequent consistent truncation to $SU(2,2|8), SU(2,2|6)$ and $SU(2,2|5)$.

  The standard supersymmetry multiplets of $\cN\geq 5$ supergravity correspond to massless doubleton representations of conformal superalgebras $SU(2,2|\cN)$ with highest spin 2. They are given in Table \ref{Table_1}, Table \ref{Table_2}, Table \ref{Table_3} for $\cN=8,6,5$ respectively.
\begin{table}[ht]
\begin{center}
\begin{tabular}{|c|c|c|c|c|}
\hline
   ${ SL(2,\mathbb{C} ) }$ & ${ E_0} $ & $ {  SU(8)}$ & ${ U(1)}$ &{ Fields}
\\ \hline
 $(0,0)$ & 1     & ${\bf 70}$   & 0 &$\phi^{[ijkl]} $
\\ \hline
 $({1 \over 2},0)$ &  ${3 \over 2} $& ${ \bf 56}$ & 1 &$\lambda^{[ijk]}_{+} \Leftrightarrow \lambda_{\alpha}^{[ijk]}$
\\ \hline
 $(0,{1 \over 2})$   & ${3 \over 2}$   &$\overline{ \bf 56}$ & -1 &$\lambda_{-[ijk]} \Leftrightarrow \lambda_{\dot{\alpha}[ijk]}$
\\ \hline
 (1,0)  & 2 & ${\bf 28}$& 2 &$F_{\mu\nu}^{+[ij]} \Leftrightarrow F_{(\alpha\beta)}^{[ij]}$
\\ \hline
 (0,1)  & 2 & $\overline{\bf 28}$& -2 &$F_{\mu\nu [ij]}^{-} \Leftrightarrow F_{(\dot{\alpha}\dot{\beta})[ij]}$
\\ \hline
 $({3 \over 2},0)$  & ${5 \over 2}$ & ${\bf 8}$& 3 &$\partial_{[ \mu}\psi_{\nu]}^{+ i} \Leftrightarrow \psi_{(\alpha\beta\gamma)}^{i}$
\\ \hline
 $(0,{3 \over 2})$  & ${5 \over 2}$ & $\bar{\bf 8}$& -3 &$\partial_{[ \mu}\psi_{\nu] i}^{-} \Leftrightarrow \psi_{(\dot{\alpha}\dot{\beta}\dot{\gamma}) i}$
\\ \hline
 $(2,0)$  &  3 & $1 $& 4 &$ R_{(\alpha\beta\gamma\delta)}$
\\ \hline
 $(0,2)$  & 3 & $1 $& -4 &$ R_{(\dot{\alpha}\dot{\beta}\dot{\gamma}\dot{\delta} )}$
\\ \hline
\end{tabular}
\medskip
\caption{\small  \label{Table_1}
The fields of linearized $\cN=8$ supergravity in four dimensions, which satisfy massless free field equations and massless representations of Poincare group. They lift uniquely to those of the conformal group. It is a CPT self-conjugate doubleton supermultiplet of $SU(2,2|8)$.  }
\end{center}
\end{table}
\begin{table}[ht]
\begin{center}
\begin{tabular}{|c|c|c|c|c|}
\hline
   ${ SL(2,\mathbb{C} ) }$ & ${ E_0} $ & $ {  SU(6)}$ & ${ U(1)}$ &{ Fields}
\\ \hline
 $(0,2)$ & 3     & ${\bf 1}$   & -2 & $R_{(\dot{\alpha}\dot{\beta}\dot{\gamma}\dot{\delta})} $
\\ \hline
 $(0,3/2)$ &  ${5/2} $& ${ \bf \bar{6}}$ & -3/2 & $\psi_{(\dot{\alpha}\dot{\beta}\dot{\gamma})i }$
\\ \hline
 $(0,1)$   & $2$   &${ \bf \bar{15}}$ & -1 &$F^-_{\mu\nu [ij]} \Leftrightarrow F_{(\dot{\alpha}\dot{\beta})[ij]}$
\\ \hline
 (0,1/2)  & 3/2 & ${\bf {20}}$& -1/2 &$\lambda^{[ijk]}_{\dot{\alpha}}$
\\ \hline
 $(0,0)$  &  1 & ${\bf 15} $& 0 & $ \phi_{[ijkl]} $
\\ \hline
 $(1/2,0)$  & 3/2 & ${\bf 6} $& 1/2 & $\lambda_{\alpha}^i$
\\ \hline
 $(1,0)$  & 2 & ${\bf{1}} $& 1 & $F_{(\alpha\beta)}$
\\ \hline
\end{tabular}
\hskip 0 cm \begin{tabular}{|c|c|c|c|c|}
\hline
   ${ SL(2,\mathbb{C} ) }$ & ${ E_0} $ & $ {  SU(6)}$ & ${ U(1)}$ &{ Fields}
\\ \hline
 $(0,1)$ & 2     & ${\bf 1}$   & -1 &$F_{(\dot{\alpha}\dot \beta)} $
\\ \hline
 $(0,1/2)$ & 3/2     & ${\bf \bar{6}}$   & -1/2 &$\lambda_{\dot{\alpha}i} $
\\ \hline
 $(0,0)$ &  ${1} $& ${ \bf \bar{15}}$ & 0 & $\phi^{[ijkl]} $
\\ \hline
 $(1/2,0)$   & ${3/2}$   &$ {\bf 20}$ & 1/2 & $\lambda^{[ijk]}_\alpha$
\\ \hline
 (1,0)  & 2 & ${\bf 15}$& 1 &$F_{\mu\nu}^{+[ij]} \Leftrightarrow F_{(\alpha\beta)}^{[ij]}$
\\ \hline
 $(3/2,0)$  &  5/2 & ${\bf 6} $& 3/2 &$ \psi_{(\alpha\beta\gamma)}^i$
\\ \hline
 $(2,0)$  & 3 & ${\bf 1} $& 2 &$ R_{(\alpha\beta\gamma\delta )}$
\\ \hline
\end{tabular}
\medskip
\caption{\small \label{Table_2}
The  irreducible chiral doubleton supermultiplet of $SU(2,2|6)$ and its conjugate, with the highest spin $2$ corresponding to the fields of $\cN=6$ supergravity.
}
\end{center}
\end{table}

\begin{table}[ht]
\begin{center}
\begin{tabular}{|c|c|c|c|c|}
\hline
   ${ SL(2,\mathbb{C} ) }$ & ${ E_0} $ & $ {  SU(5)}$ & ${ U(1)}$ &{ Fields}
\\ \hline
 $(0,2)$ & 3     & ${\bf 1}$   & -2 & $R_{(\dot{\alpha}\dot{\beta}\dot{\gamma}\dot{\delta})} $
\\ \hline
 $(0,3/2)$ &  ${5/2} $& ${ \bf \bar{5}}$ & -3/2 & $\psi_{(\dot{\alpha}\dot{\beta}\dot{\gamma})i}$
\\ \hline
 $(0,1)$   & $2$   &${ \bf \bar{10}}$ & -1 &$F^-_{\mu\nu[ij]} \Leftrightarrow F_{(\dot{\alpha}\dot{\beta})[ij]}$
\\ \hline
 (0,1/2)  & 3/2 & ${\bf 10}$& -1/2 &$\psi_{\dot{\alpha}[ijk]}$
\\ \hline
 $(0,0)$  &  1 & ${\bf 5} $& 0 & $ \phi_{[ijkl]} $
\\ \hline
 $(1/2,0)$  & 3/2 & ${\bf 1} $& 1/2 & $\psi_{\alpha}$
\\ \hline
\end{tabular}
\hskip 0 cm \begin{tabular}{|c|c|c|c|c|}
\hline
   ${ SL(2,\mathbb{C} ) }$ & ${ E_0} $ & $ {  SU(5)}$ & ${ U(1)}$ &{ Fields}
\\ \hline
 $(0,1/2)$ & 3/2     & ${\bf 1}$   & -1/2 &$\psi_{\dot{\alpha}} $
\\ \hline
 $(0,0)$ &  ${1} $& ${ \bf \bar{5}}$ & 0 & $\phi^{[ijkl]} $
\\ \hline
 $(1/2,0)$   & ${3/2}$   &${ \bf \bar{10}}$ & 1/2 & $\psi^{[ijk]}_\alpha$
\\ \hline
 (1,0)  & 2 & ${\bf 10}$& 1 &$F_{\mu\nu}^{+[ij]} \Leftrightarrow F_{(\alpha\beta)}^{[ij]}$
\\ \hline
 $(3/2,0)$  &  5/2 & ${\bf 5} $& 3/2 &$ \psi_{(\alpha\beta\gamma)}^i$
\\ \hline
 $(2,0)$  & 3 & ${\bf 1} $& 2 &$ R_{(\alpha\beta\gamma\delta )}$
\\ \hline
\end{tabular}
\medskip
\caption{\small \label{Table_3}
The  irreducible chiral doubleton supermultiplet of $SU(2,2|5)$ and its conjugate corresponding to the $\cN=5$ supergravity multiplet. 
 }
\end{center}
\end{table}

All these standard multiplets have vectors  and scalars of classical supergravity. Tables \ref{Table_1}, Table \ref{Table_2}, Table \ref{Table_3} for $\cN=8,6,5$ respectively are in one-to-one correspondence with the linearized conformal superfields presented in \cite{Freedman:2017zgq}. The linearized candidate counterterms  \cite{Kallosh:1980fi, Freedman:2017zgq}  are constructed using these supermultiplets.

One way to double the set of vector fields is simply to take two sets of the graviton supermultiplets above. However, this is not satisfactory for the BN proposal since the numbers of scalars is doubled and one would have to couple the graviton supermultiplets to themselves which is known not to be possible.  We need to study  all possible supermultiplets in $SU(2,2|\cN)$ with vectors in the same representations as in the classical case. To make sure that we check all options we proceed with the superoscillator construction.


\section{ All {\boldmath$SU(2,2|\cN)$} supermultiplets with vectors in anti-symmetric tensor representations of $SU(\cN)$ for {\boldmath $\cN=5,6,8$}}
The oscillator construction of the unitary supermultiplets of extended superconformal algebras in four dimensions  were studied in  \cite{Gunaydin:1984fk,Gunaydin:1984vz,Gunaydin:1998sw,Gunaydin:1998jc}. As has been proven recently the oscillator method yields all the unitary representations of the Lie superalgebras of the form $SU(m,n|p+q)$ \cite{Gunaydin:2017lhg}.
The superalgebra $SU(2,2|p+q)$ has a three graded decomposition with
respect to its compact subsuperalgebra $SU(2|p)\times SU(2|q) \times U(1)$
\eq
g = L^{+} \oplus L^{0} \oplus L^{-} \ ,
\en
\bea
[L^{0},L^{\pm}] = L^{\pm} \ ,  \qquad
[L^{+},L^{-}] =L^{0} \ , \qquad
[L^{+},L^{+}] = 0=[L^{-},L^{-}] \ .
\eea
Here $L^{0}$ represents the generators of
$SU(2|p) \times SU(2|q) \times U(1)$.
The Lie superalgebra $SU(2,2|p+q)$
 can be realized in terms of bilinear combinations of bosonic and
fermionic annihilation and creation operators $\xi_{A}$
($\xi^{A}={\xi_{A}}^{\dagger}$) and $\eta_{M}$
($\eta^{M}={\eta_{M}}^{\dagger}$)
which transform covariantly and contravariantly
under the  $SU(2|p)$ and $SU(2|q)$ subsuperalgebras of $SU(2,2|p+q)$

\be
\xi_{A} = \left( \begin{array}{c} a_{\alpha} \\
                        \alpha_{x}   \end{array}  \right) ,\quad
\xi^{A}=\left( \begin{array}{c} a^{\alpha} \\
                        \alpha^{x}   \end{array} \right) \ ,
\ee
\bea
\eta_{M} = \left(\begin{array}{c} b_{\dot \alpha} \\
                        \beta_{i}  \end{array} \right) , \quad
\eta^{M} = \left(\begin{array}{c} b^{\dot \alpha} \\
                        \beta^{i} \end{array} \right) \ ,
\eea
with $\alpha, \beta=1,2$; $x,y=1,2,..,p$; $\dot \alpha, \dot \beta=1,2$; $i,j=1,2,..,q$ and

\eq
[a_\alpha, a^\beta] = \delta_{\alpha}^{\beta} , \quad
\{\alpha_{x}, \alpha^{y}\} = \delta_{x}^{y} \ ,
\en
\eq
[b_{\dot \alpha}, b^{\dot \beta}] = \delta_{\dot \alpha}^{\dot \beta} , \quad
\{\beta_{i}, \beta^{j}\} = \delta_{i}^{j} \ .
\en
The generators of $SU(2,2|p+q)$ are given in terms of the above
superoscillators as
\bea
L^{-} &=& {\vec{\xi}}_{A} \cdot {\vec{\eta}}_{M} \cr
L^{0} &=& {\vec{\xi}}^{A} \cdot {\vec{\xi}}_{B}
\oplus {\vec{\eta}}^{M} \cdot {\vec{\eta}}_{N} \cr
L^{+} &=& {\vec{\xi}}^{A} \cdot {\vec{\eta}}^{M}
\eea
and we refer to \cite{Gunaydin:1984fk,Gunaydin:1984vz,Gunaydin:1998sw,Gunaydin:1998jc} for details. Massless conformal supermultiplets are obtained by taking one set ( color)  of super-oscillators which are referred to as doubleton supermultiplets.

The unitary irreducible representations of $SU(2,2 |p+q)$ algebra are constructed over the super Fock space of these oscillators. One chooses a set of states $|\Omega\rangle $, \footnote{ By an abuse of notation  this set of states will be  referred to as the ``ground state" or  as the ``lowest weight vector". They correspond simply to the lowest ``energy" irrep of the compact subsuperalgebra $SU(2|p)\oplus SU(2|q)$ if one identifies  the energy operator with the generator of $U(1)$ that defines the 3-grading.} that are annihilated by the ``lowering'' operators
$L^-$ and  transform irreducibly under the grade zero subalgebra $SU(2|p)\oplus SU(2|q) \oplus U(1)$. Then by repeated application of the raising operators
$L^{+}$ one generates an infinite set of states that form the basis of a unitary irreducible representation of $SU(2,2|p+q)$:
\eq
 |\Omega \rangle ,\quad  L^{+1}|\Omega \rangle ,\quad
L^{+1} L^{+1}|\Omega \rangle , ...
\en
Possible lowest weight vectors of massless conformal supermultiplets in $d=4$ are either of the form
\eq
|\Omega \rangle = \xi^{A_1} \xi^{A_2}...\xi^{A_{P}} |0\rangle
= |\underbrace{\sgenrowbox}_{P}, 1 \rangle
\label{general}\en
or the form
\eq
|\Omega \rangle = \eta^{A_1} \eta^{A_2}...\eta^{A_{Q}} |0\rangle
= |1, \underbrace{\sgenrowbox}_{Q} \rangle
\label{generalC}\en
Here we use boxes with slashes for the {\it super-Young-tableaux} and $P$ and $Q$ are arbitrary integers. We should note that one obtains the same set of representations of $SU(2,2|\cN)$ irrespective of the choice of $p$ and $q$ so long as
$p+q=\cN$ \footnote{ This is true not only for the massless conformal supermultiplets but for all the unitary supermultiplets of
$SU(2,2|p+q)$ \cite{Gunaydin:2017lhg}.}.

We find it convenient for our purpose to use the basis where $p=0$ and $q=\cN$ with $\alpha, \beta=1,2$;  $\dot \alpha, \dot \beta=1,2$; $i,j=1,2,..,\cN$ \footnote{Equivalently one can choose a basis with $q=0$ and $p=\cN$.}

\be
\xi_{A} = \left( \begin{array}{c} a_{\alpha} \\
                        0  \end{array}  \right) ,\quad
\xi^{A}=\left( \begin{array}{c} a^{\alpha} \\
                       0   \end{array} \right) ,
\ee
\bea
\eta_{M} = \left(\begin{array}{c} b_{\dot \alpha} \\
                        \beta_{i}  \end{array} \right) , \quad
\eta^{M} = \left(\begin{array}{c} b^{\dot \alpha} \\
                        \beta^{i} \end{array} \right) .
\eea

Just looking at eq. \rf{general} we deduce that  that scalars must be $U(\cN)$ singlets. In \rf{generalC} where for scalars we need an anti-symmetric set of $\cN$ operators $\beta^i$, which for all $\cN$ form a singlet. This  gives us a prediction that all supermultiplets for $\cN\geq 5$ with required vectors, other than the graviton supermultiplets,  have scalars that are singlets of $SU(\cN)$, according to the formulas \rf{general} and   \rf{generalC}. We confirm this prediction by constructing the corresponding supermultiplets explicitly.

Below we present detailed form of the supermultiplets  for each $\cN$ with vectors in anti-symmetric tensor representation of $SU(\cN)$.  $\cN=6$ case is special in that the graviton supermultiplet has, in addition,  a vector that is a singlet of $SU(6)$.

In $\cN=5$  we have a vector in ${ \bf 10}$ and ${\bf \bar{10}}$ of $SU(5)$, whereas the scalar is a singlet in $SU(5)$. In  Table \ref{Table_4} we give the supermultiplets whose
 lowest weight vectors are $|1,1\rangle =|0\rangle $ and , $|1,\underbrace{\sgenrowbox}_{5} \rangle $, respectively.
\begin{table}[ht]
\begin{center}
\begin{tabular}{|c|c|c|}
\hline
   ${ SL(2,\mathbb{C} ) }$ &  $ {  SU(5)}$ & ${ U(1)}$
\\ \hline
 $(0,0)$ &  ${\bf 1}$   & 0
\\ \hline
 $({1 /2},0)$ &   ${ \bf 5}$ & 1/2
\\ \hline
 $(1,0)$   & ${ \bf 10}$ & 1
\\ \hline
 $({3/ 2},0)$  &  ${\bf \bar{10}}$ & 3/2
\\ \hline
 $(2,0)$  &   ${\bf \bar{5}} $& 2
\\ \hline
 $(5/2,0)$  &  ${\bf \bar{1}} $& 5/2
\\ \hline
\end{tabular}
\hskip 1 cm \begin{tabular}{|c|c|c|}
\hline
  ${ SL(2,\mathbb{C} ) }$ &  $ {  SU(5)}$ & ${ U(1)}$
\\ \hline
 $(0,5/2)$ &  ${\bf 1}$   & -5/2
\\ \hline
 $(0,2)$ &  ${ \bf 5}$ & -2 \\ \hline
 $(0,3/2)$   & ${ \bf 10}$ & -3/2
\\ \hline
 (0,1)  &  ${\bf \bar{10}}$& -1
\\ \hline
 $(0,1/2)$  &   ${\bf \bar{5}} $& -1/2
\\ \hline
 $(0,0)$  &  ${\bf \bar{1}} $& 0
\\ \hline
\end{tabular}
\medskip
 \caption{\label{Table_4}
The  irreducible chiral doubleton supermultiplets of $SU(2,2|5)$ and its conjugate, with the highest spin $5/2$.  The scalars here are singlets of $SU(5)$. The vector field strengths and their duals transform in ${\bf 10}$ and ${\bf \bar{10}}$ of $SU(5)$.}
\end{center}
\end{table}

In $\cN=6$
 we have a vector in ${ \bf 15}$ and ${\bf \bar{15}}$ of $SU(6)$, whereas the scalar is a singlet in $SU(6)$. In  Table \ref{Table_5} we give the supermultiplets of $SU(2,2|6)$ whose
 lowest weight vectors are $|1,1\rangle =|0\rangle $ and , $|1,\underbrace{\sgenrowbox}_{6} \rangle $, respectively.
\begin{table}[ht]
\begin{center}
\begin{tabular}{|c|c|c|}
\hline
   ${ SL(2,\mathbb{C} ) }$ &  $ {  SU(6)}$  &{ $U(1)$}
\\ \hline
 $(0,0)$ & ${\bf 1}$   & 0
\\ \hline
 $(1/2,0)$ & $ { \bf 6}$ & 1/2
\\ \hline
 $(1,0)$ & ${ {\bf 15}}$ & 1 \\ \hline
  $(3/2,0)$ & $  { \bf 20}$ & 3/2\\ \hline
  $(2,0)$ & $ {\overline{ \bf 15}}$ & 2\\ \hline
 $(5/2,0)$   &${\overline {\bf 6}}$ & 5/2
\\ \hline
 (3,0)  &  ${\overline{\bf 1}}$&3\\ \hline
\end{tabular}
\hskip 1.5 cm \begin{tabular}{|c|c|c|}
\hline
   ${ SL(2,\mathbb{C} ) }$ &  $ {  SU(6)}$ & $U(1)$
\\ \hline
 $(0,3)$ & ${ \bf 1}$ & -3 \\ \hline
  $(0,5/2)$ & ${ \bf 6}$ & -5/2\\ \hline
  $(0,2)$ & ${ \bf 15}$ & -2\\ \hline
 $(0,3/2)$   &$\overline{ \bf 20}$ & -3/2
\\ \hline
 (0,1)  &  $\overline{{\bf 15}}$&-1
\\ \hline
 (0,1/2)  & $\overline{\bf 6}$& -1/2\\ \hline
  $(0,0)$  &  $\overline{{\bf 1}}$& 0\\ \hline
\end{tabular}
\medskip
\caption{\small \label{Table_5}
The  irreducible chiral doubleton supermultiplets of $SU(2,2|6)$ and its conjugate, with the highest spin $3$. Scalars here are singlets of $SU(6)$. The vector field strengths and their duals transform in ${\bf 15}$ and ${\bf \bar{15}}$ of $SU(6)$.
 }
\end{center}
\end{table}

In $\cN=8$  we have vector field strengths and their duals  in ${ \bf 28}$ and ${\bf \bar{28}}$ of $SU(8)$, whereas the scalar is a singlet in $SU(8)$. In  Table \ref{Table_6} we give the supermultiplets of $SU(2,2|8)$ whose
 lowest weight vectors are $|1,1\rangle =|0\rangle $ and , $|1,\underbrace{\sgenrowbox}_{8} \rangle $, respectively.
\begin{table}[ht]
\begin{center}
\begin{tabular}{|c|c|c|}
\hline
   ${ SL(2,\mathbb{C} ) }$ &  $ {  SU(8)}$  &{ $U(1)$}
\\ \hline
 $(0,0)$ & ${\bf 1}$   & 0
\\ \hline
 $(1/2,0)$ & $ { \bf 8}$ & 1/2
\\ \hline
 $(1,0)$ & ${ {\bf 28}}$ & 1 \\ \hline
  $(3/2,0)$ & ${  { \bf 56}}$ & 3/2\\ \hline
  $(2,0)$ & ${ \bf 70}$ & 2\\ \hline
 $(5/2,0)$   &${\overline {\bf 56}}$ & 5/2
\\ \hline
 (3,0)  &  ${\overline{\bf 28}}$&3\\ \hline
 (7/2,0)  & ${\bf 8}$& 7/2\\ \hline
  $(4,0)$  &  ${\overline {\bf 1}}$& 4
\\ \hline
\end{tabular}
\hskip 1.5 cm \begin{tabular}{|c|c|c|}
\hline
   ${ SL(2,\mathbb{C} ) }$ &  $ {  SU(8)}$ & $U(1)$
\\ \hline
 $(0,4)$ & ${\bf 1}$   & -4\\ \hline
 $(0,7/2)$ & ${ \bf 8}$ & -7/2
\\ \hline
 $(0,3)$ & ${ \bf 28}$ & -3 \\ \hline
  $(0,5/2)$ & ${ \bf 56}$ & -5/2\\ \hline
  $0,2)$ & ${ \bf 70}$ & -2\\ \hline
 $(0,3/2)$   &$\overline{ \bf 56}$ & -3/2
\\ \hline
 (0,1)  &  $\overline{{\bf 28}}$&-1
\\ \hline
 (0,1/2)  & $\overline{\bf 8}$& -1/2\\ \hline
  $(0,0)$  &  $\overline{{\bf 1}}$& 0\\ \hline
\end{tabular}
\medskip
\caption{\small \label{Table_6}
The  irreducible chiral doubleton supermultiplets of $SU(2,2|8)$ and its conjugate, with the highest spin $4$. Scalars here are singlets of $SU(8)$.  The vector field strengths and their duals transform in ${\bf 28}$ and ${\bf \bar{28}}$ of $SU(8)$.
 }
\end{center}
\end{table}

Finally, in Table \ref{Table_7}  we show the $\cN=6$ multiplets with an $SU(6)$ singlet vector, without scalars. Their lowest weight  vectors are $a^\alpha a^\beta|0\rangle $ and  $|1,\underbrace{\sgenrowbox}_{8} \rangle $, respectively. This is a special property of $\cN=6$ supergravity where 16 vector field strengths transform in ${ \bf 15}$  of $SU(6)$ plus a  singlet. This is not given by the general formulas  \rf{general}, \rf{generalC}.

\begin{table}[ht]
\begin{center}
\begin{tabular}{|c|c|c|}
\hline
   ${ SL(2,\mathbb{C} ) }$ &  $ {  SU(6)}$  &{ $U(1)$}
\\ \hline
 $(1,0)$ & ${ {\bf 1}}$ & 1 \\ \hline
  $(3/2,0)$ & ${  { \bf 6}}$ & 3/2\\ \hline
  $(2,0)$ & ${ \bf 15}$ & 2\\ \hline
 $(5/2,0)$   &$ {\bf 20}$ & 5/2
\\ \hline
 (3,0)  &  ${\overline{\bf 15}}$&3\\ \hline
 (7/2,0)  & ${\overline {\bf 6}}$& 7/2\\ \hline
  $(4,0)$  &  ${\overline {\bf 1}}$& 4
\\ \hline
\end{tabular}
\hskip 1.5 cm \begin{tabular}{|c|c|c|}
\hline
   ${ SL(2,\mathbb{C} ) }$ &  $ {  SU(8)}$ & $U(1)$
\\ \hline
 $(0,4)$ & ${\bf 1}$   & -4\\ \hline
 $(0,7/2)$ & ${ \bf 6}$ & -7/2
\\ \hline
 $(0,3)$ & ${ \bf 15}$ & -3 \\ \hline
  $(0,5/2)$ & ${ \bf 20}$ & -5/2\\ \hline
  $0,2)$ & $\overline{ \bf 15}$ & -2\\ \hline
 $(0,3/2)$   &$\overline{ \bf 6}$ & -3/2
\\ \hline
 (0,1)  &  $\overline{{\bf 1}}$&-1
\\ \hline
\end{tabular}
\medskip
\caption{\small \label{Table_7}
The  irreducible chiral doubleton supermultiplets of $SU(2,2|6)$ and its conjugate, with the highest spin $4$. Vectors here are singlets of $SU(6)$ and scalars are absent.
 }
\end{center}
\end{table}

Note that in Tables \ref{Table_4} - \ref{Table_7}  the  supermultiplets break the spin 2 barrier, $s>2$, maximum spin  is  $4$, and contain    multiple  gravitons.  Therefore we see that, in addition to the fact that the scalars in these supermultiplets are not consistent with  $E7$ type duality,  they all contain higher spin fields and multiple gravitons which add further support to the arguments  that the source of deformation with properties required by the proposal of   \cite{Bossard:2011ij} are not available.



\section{  {\boldmath$SU(2,2|\cN +n)$}  supermultiplets containing vectors in anti-symmetric tensor representations of $SU(\cN)$ for {\boldmath $\cN=5,6,8$} }

The minimal CPT self-conjugate unitary supermultiplet that contains two sets of vector fields transforming in the 28 of $SU(8)$ is the doubleton supermultiplet of $SU(2,2|10)$ which we  already discussed in \cite{Gunaydin:2013pma}. Here we  show it  in  Table \ref{Table_GK}. We consider a decomposition of $SU(10)$ under the $SU(8)\times SU(2)$. In the last column of Table \ref{Table_GK} we show this decomposition. We have here the double set of 28 vectors, we have $(\overline{28},2)$ and $({28},2)$. Note, however, that the scalars here are also in $(70,2)$, i. e. twice the amount we need.  This is not accidental, if one truncates this unitary supermultiplet of $SU(2,2|10)$ by  throwing out all the $SU(2)$ singlet states one gets two copies of the  CPT self-conjugate supermultiplet of $SU(2,2|8)$. Therefore it is not valid for a source of deformation.

The CPT self-conjugate unitary supermultiplet of $SU(2,2|8+2n)$ for $n>0$ contains $\frac{(2n)!}{n!}$ pairs of vector field strength multiplets transforming in $(28+\overline{28})$ of $SU(8)$ subgroup. But also the number of scalars is increased, so all these models are not working.

\begin{table}[ht]
\begin{center}
\begin{tabular}{|c|c|c|c|c|c|}
\hline
   ${ SL(2,\mathbb{C} ) }$ & ${ E_0} $ & $ {  SU(10)}$ & ${ U(1)}$ &{ Fields}
& $SU(8)\times SU(2)$ \\ \hline
 $(0,0)$ & 1     & ${\bf 252}$   & 0 &$\phi^{[ijklm]} $ &$(56,1)+(\overline{56},1)+ (70,2)$
\\ \hline
 $({1 \over 2},0)$ &  ${3 \over 2} $& ${ \bf 210}$ & 1 &$\lambda^{[ijkl]}_{+} \equiv \lambda_{\alpha}^{[ijkl]}$ & $(70,1) +(28,1)+(56,2)$
\\ \hline
 $(0,{1 \over 2})$   & ${3 \over 2}$   &$\overline{ \bf 210}$ & -1 &$\lambda_{-[ijkl]} \equiv \lambda_{\dot{\alpha}[ijkl]}$ &$(70,1) +(\overline{28},1)+(\overline{56},2)$
\\ \hline
 (1,0)  & 2 & ${\bf 120}$& 2 &$F_{\mu\nu}^{+[ijk]} \equiv F_{(\alpha\beta)}^{[ijk]}$ &$(56,1)+(8,1)+(28,2)$
\\ \hline
 (0,1)  & 2 & $\overline{\bf 120}$& -2 &$F_{\mu\nu [ijk]}^{-} \equiv F_{(\dot{\alpha}\dot{\beta})[ijk]}$ &$(\overline{56},1)+(\overline{8},1)+(\overline{28},2)$
\\ \hline
 $({3 \over 2},0)$  & ${5 \over 2}$ & ${\bf 45}$& 3 &$\partial_{[ \mu}\psi_{\nu]}^{+ [ij]} \equiv \psi_{(\alpha\beta\gamma)}^{[ij]}$ & $(28,1)+(1,1)+(8,2)$
\\ \hline
 $(0,{3 \over 2})$  & ${5 \over 2}$ & $\bar{\bf 45}$& -3 &$\partial_{[ \mu}\psi_{\nu] [ij]}^{-} \equiv \psi_{(\dot{\alpha}\dot{\beta}\dot{\gamma}) [ij]}$ &$(\overline{28},1)+(1,1)+(\overline{8},2)$
\\ \hline
 $(2,0)$  &  3 & ${\bf10} $& 4 &$ R^i_{(\alpha\beta\gamma\delta)}$ & $(8,1)+(1,2)$
\\ \hline
 $(0,2)$  & 3 & $\overline{\bf 10} $& -4 &$ R_{(\dot{\alpha}\dot{\beta}\dot{\gamma}\dot{\delta}) i}$ &$(\overline{8},1)+(1,2)$
\\ \hline
$({5 \over 2},0)$ & ${7 \over 2}$& 1 & 5 & $ R_{(\alpha\beta\gamma\delta\epsilon)}$ & $(1,1)$ \\ \hline
$(0, {5 \over 2})$ & ${7 \over 2}$& 1 & 5 & $ R_{(\dot{\alpha}\dot{\beta}\dot{\gamma}\dot{\delta}\dot{\epsilon})}$ &$(1,1)$ \\ \hline
\end{tabular}
\medskip
\caption{\small \label{Table_GK}
An example of the CPT-self-conjugate doubleton supermultiplet of $SU(2,2|10)$.    $i,j,k,..=1,2,..,10 $ are the $SU(10)$ $R$-symmetry  indices.
}
\end{center}
\end{table}

To proceed with a more general case and to incorporate into our analysis both $\cN=5$ and $\cN=6$ we would like to take into account the subtlety with $\cN=6$ case, which we already discussed before. Namely, in all $\cN\geq 5$ supergravities the vector field strength transform as an antisymmetric tensor $F^{+[ij]}_{\mu\nu}$, $F^{-}_{[ij] \mu\nu}$. In $\cN=6$ case we also have singlets of the $R$-symmetry group. One way to get a doublet of vector fields is to increase the $R$-symmetry from $SU(\cN)$ to $SU(\cN+2)$ and choose supermultiplets in which the vector field strengths transform as
\be
F^{+[IJK]}_{\mu\nu} \ ,  \qquad F^{-}_{[IJK] \mu\nu} \ , \qquad \qquad I,J= 1, \cdots \cN+2 \ .
\ee
Under restriction to $SU(\cN)$ we then have $SU(\cN+2) \supset SU(\cN)\times SU(2) \times U(1)$ and we get the following decomposition
\bea
{(\cN+2 ) (\cN+1) (\cN)\over 3!}&=& \Big( {(\cN ) (\cN-1) (\cN-1)\over 3!}, 1\Big) + \Big( {(\cN ) (\cN-1) \over 2!}, 2\Big) + (\cN, 1)\cr
\cr
F^{+[IJK]}_{\mu\nu} &= & \hskip 1 cmF^{+[ijk]}_{\mu\nu}  \hskip 2.2 cm + F^{+[ij]a}_{\mu\nu}  \hskip 1.5 cm+ F^{+ i[ab]}_{\mu\nu}
\eea
In the case of $\cN=8$ $SU(2,2|10)$ which we show in the Table \ref{Table_GK}, we see that indeed, we have a doublet of required vectors. However, the scalars in ${\bf 252}$ decompose as $(56,1) + (\overline{56},1) + (70, 2)$ under $SU(8) \times SU(2)$ ! So we have a double amount of scalars compared to the required ones.

In general, to obtain the states corresponding to fields strengths $F^{+[IJK]}_{\mu\nu}$, transforming in the $(1,0)$ representation of the Lorentz group, one needs  specific supermultiplets. Its highest helicity state must either be  $(0,5/2)$ in a singlet of $SU(\cN+2)$ or its lowest helicity state must be $(1/2, 0)$ in a singlet of $SU(\cN+2)$. The first one leads to scalars transforming as an anti-symmetric tensor of rank 5 under $SU(\cN+2)$.

$$SU(\cN+2) \supset SU(\cN)\times SU(2) \times U(1)$$

\be
\phi^{[IJKLM]}= \phi^{[ijklm]}+ \phi^{[ijkl]a}+ \phi^{[ijk][ab]}
\ee
\bea
&252 = (\overline {56}, 1) + (70,2) + (56,1) \qquad  \cN=8\cr
\cr
&\overline {126} = (\overline {21}, 1) + (\overline {35},2) + (35,1) \qquad  \cN=7\cr
\cr
&\overline {56} = (\overline {6}, 1) + \, (\overline {15},2) + (20,1) \qquad  \cN=6\cr
\cr
&\overline {21} = (1, 1) + \, (\overline {5},2) + \, (\overline {10},1) \qquad  \cN=5
\eea

In the second case for the supermultiplet with the lowest helicity state $(1/2, 0)$ in a singlet of $SU(\cN+2)$ we find that the scalars transform in the fundamental representation of $SU(\cN+2)$ which decomposes as

$$SU(\cN+2) \supset SU(\cN)\times SU(2) $$
\bea
&10 = (8, 1) + (70,2)  \qquad  \cN=8\cr
\cr
&9 = (7, 1) + (1,2)  \qquad  \cN=7\cr
\cr
&8= ( {6}, 1) + \, (1,2)  \qquad  \cN=6\cr
\cr
&7 = (5, 1) + \, (1,2)  \qquad  \cN=5
\eea
Only for $\cN=5$ the scalars in $(5,1)$ are in the right representation. However, in all cases above we have unwanted doublets of scalars.
Thus, here again, we checked all possibilities of a supersymmetric doubling of vectors. We always see that it is impossible to keep the required scalars.

Besides, all new supermultiplets involved have higher spins with $s>2$ and all issues discussed in \cite{Gunaydin:2013pma} for $\cN=8$ persist for $\cN=5,6$, namely the presence of a finite number of states with higher spins $s>2$ and presence of many fields with $s=2$. No known interacting theories exist that describe the coupling of a finite number of higher spin fields to gravity. Furthermore the question of consistent coupling between a finite number of massless $s=2$ fields has been investigated in \cite{Boulanger:2000rq} where the inconsistency of interacting multi-graviton theories was established.  In addition as we have emphasized above  possible sources of deformation introduce   additional scalar fields beyond the original coset. Therefore we conclude that BN proposal is incompatible with linearized on-shell supersymmetry.

\section{Superfields}
We have discussed candidate supermultiplets including vector duality doublet so far. Superfield expression of such multiplets makes the structure more transparent, and therefore, we show such superfields corresponding to multiplets in Tables \ref{Table_4}-\ref{Table_6} and one originating from a larger supermultiplet in Table \ref{Table_GK}. We are using the notation of  \cite{Freedman:2017zgq,Freedman:2018mrv}.
On-shell supermultiplets we have  studied above can be simply expressed as  chiral superfields. For $\cN=5$, the supermultiplet in Table \ref{Table_4} is given simply by the superfield
\begin{align}
\Phi(y,\theta)=&\phi+\theta_i^\a\chi^i_\a+\theta_i^\a\theta_j^\b F_{\a\b}^{ij}+\theta_i^\a\theta_j^\b\theta_k^\g\epsilon^{ijklm}\psi_{lm\a\b\g}\nonumber\\
&+\theta_i^\a\theta_j^\b\theta_k^\g\theta_l^\delta\epsilon^{ijklm}C_{m\a\b\g\delta}+\theta_i^\a\theta_j^\b\theta_k^\g\theta_l^\delta\theta_m^{\epsilon}\epsilon^{ijklm}E_{\a\b\g\delta\epsilon} \ .
\end{align}
For the supermultiplet of $\cN=6$ in Table \ref{Table_5}, we find
\begin{align}
\Phi(y,\theta)=&\phi+\theta_i^\a\chi^i_\a+\theta_i^\a\theta_j^\b F_{\a\b}^{ij}+\theta_i^\a\theta_j^\b\theta_k^\g\epsilon^{ijklmn}\psi_{lmn\a\b\g}+\theta_i^\a\theta_j^\b\theta_k^\g\theta_l^\delta\epsilon^{ijklmn}C_{mn\a\b\g\delta}\nonumber\\
&+\theta_i^\a\theta_j^\b\theta_k^\g\theta_l^\delta\theta_m^{\epsilon}\epsilon^{ijklmn}E_{n\a\b\g\delta\epsilon}+\theta_i^\a\theta_j^\b\theta_k^\g\theta_l^\delta\theta_m^{\epsilon}\theta_n^{\theta}\epsilon^{ijklmn}G_{\a\b\g\delta\epsilon\theta} \ .
\end{align}
For $\cN=8$, we found two candidate supermultiplets: the first one corresponds to  the scalar superfield:
\begin{align}
\Phi(y,\theta)=&\phi+\theta_i^\a\chi^i_\a+\theta_i^\a\theta_j^\b F_{\a\b}^{ij}+\theta_i^\a\theta_j^\b\theta_k^\g\epsilon^{ijklmnpq}\psi_{lmnpq\a\b\g}+\theta_i^\a\theta_j^\b\theta_k^\g\theta_l^\delta\epsilon^{ijklmnpq}C_{mn\a\b\g\delta}\nonumber\\
&+\theta_i^\a\theta_j^\b\theta_k^\g\theta_l^\delta\theta_m^{\epsilon}\epsilon^{ijklmnpq}E_{npq\a\b\g\delta\epsilon}+\theta_i^\a\theta_j^\b\theta_k^\g\theta_l^\delta\theta_m^{\epsilon}\theta_n^{\theta}\epsilon^{ijklmnpq}G_{pq\a\b\g\delta\epsilon\theta}\nonumber\\
&+\theta_i^\a\theta_j^\b\theta_k^\g\theta_l^\delta\theta_m^{\epsilon}\theta_n^{\theta}\theta_p^{\zeta}\epsilon^{ijklmnpq}H_{q\a\b\g\delta\epsilon\theta\zeta}+\theta_i^\a\theta_j^\b\theta_k^\g\theta_l^\delta\theta_m^{\epsilon}\theta_n^{\theta}\theta_p^{\zeta}\theta_q^\eta \epsilon^{ijklmnpq}J_{\a\b\g\delta\epsilon\theta\zeta\eta}
\end{align}
Note that, for each component field, the Lorentz indices $\a,\b,\cdots, \eta$ are completely symmetrized, e.g. $J_{\a\b\g\delta\epsilon\theta\zeta\eta}=J_{(\a\b\g\delta\epsilon\theta\zeta\eta)}$. These scalar superfields in $\cN\geq5$ could be candidates for supermultiplets providing extra (dual) vector fields. As we see, however, each of the superfields contains scalar fields which are duality singlets, and also has fields with spin $s\geq2$. Therefore, we can not use such superfields to implement the deformation.

There is also another possibility to have vector doublet, which originates from a truncation of $\cN=10$ multiplet for $\cN=8$ case,
\begin{align}
\bar{C}_{\dot\a\dot\b\dot\g\dot\delta}^a(y, \theta)&= \bar C_{\dot\a\dot\b\dot\g\dot\delta}^a(y) +\theta^{\a}_i  \partial_{\a (\dot{\a} } \bar{\psi}^{ia}_{\dot \b \dot \g \dot \delta)}
+\ft12\theta^{\a}_i \theta^{\b}_j \partial_{\a (\dot \a} \partial_{\b \dot\b} \bar{M}^{jia}_{\dot \g \dot\delta)}
+\ft1{3!} \theta^{\a}_i \theta^{\b}_j \theta^{\g}_k \partial_{\a (\dot\a} \partial_{\b \dot\b}  \partial_{\g \dot \g} \bar{\chi}_{\dot \delta)}^{kjia} \nonumber \\
&+ \ft1{4!}\theta^{\a}_i \theta^{\b}_j \theta^{\g}_k \theta^{\delta}_{\ell } \partial_{\a \dot\a} \partial_{\b \dot\b}  \partial_{\g \dot \g} \partial_{\delta \dot \delta} \phi^{\ell k jia}
+\ft1{5!3!} \theta^{\a}_i \theta^{\b}_j \theta^{\g}_k \theta^{\delta}_{\ell } \theta^{\eps}_{m} \partial_{\a \dot\a} \partial_{\b \dot\b}  \partial_{\g \dot \g} \partial_{\delta \dot \delta} \chi_{\eps npq}^a \varepsilon ^{ijk\ell mnpq} \nonumber \\
&+\ft1{6!2} \theta^{\a}_i \theta^{\b}_j \theta^{\g}_k \theta^{\delta}_{\ell } \theta^{\eps}_{m} \theta^{\zeta}_{n} \partial_{\a \dot \a} \partial_{\b \dot \b} \partial_{\g \dot \g} \partial_{\delta \dot \delta} M_{\eps \zeta pq}^a \varepsilon ^{ijk\ell mnpq} \nonumber\\
&+ \ft1{7!}  \theta^{\a}_i \theta^{\b}_j \theta^{\g}_k \theta^{\delta}_{\ell } \theta^{\kappa _1}_{m} \theta^{\kappa _2}_{n} \theta^{\kappa _3}_{p}
\partial_{\a \dot \a} \partial_{\b \dot \b} \partial_{\g \dot \g} \partial_{\delta \dot \delta} \psi_{\kappa_1\kappa _2\kappa _3 q}^a \varepsilon ^{ijk\ell mnpq} \nonumber \\&
+\ft1{8!} \theta^{\a}_i \theta^{\b}_j \theta^{\g}_k \theta^{\delta}_{\ell } \theta^{\kappa _1}_{m} \theta^{\kappa _2}_{n} \theta^{\kappa _3}_{p}\theta ^{\kappa _4}_q
  \partial_{\a \dot \a} \partial_{\b \dot \b} \partial_{\g \dot \g} \partial_{\delta \dot \delta} C_{\kappa_1\kappa _2\kappa _3\kappa _4}^a \varepsilon ^{ijk\ell mnpq}\, ,
\end{align}
where $a=1,2$ is an extra SU(2) index. In this case, the vector is doubled, but also all other components are. This leads e.g. to two
$\bf 70$ for scalars, two
gravitons, etc. and therefore, we cannot use this superfield for deformation either. One finds analogous results  with $\cN=8$ supermultiplet truncation to $\cN=6$ and $\cN=7$ supermultiplet truncation to $\cN=5$.

\section{Discussion}
The complicated situation with UV divergences in perturbative $\cN\geq 5$ supergravity \footnote{It was argued in  \cite{Kallosh:2008mq}  that the supersymmetric counterterms of $\cN=8$ supergravity are absent in the light-cone formalism.  The corresponding counterterms in the light-cone formalism were never constructed during the last decade, in agreement with the arguments in \cite{Kallosh:2008mq}. By a consistent reduction of supersymmetry, one would expect also the absence of the light-cone candidate counterterms in $\cN=6,5$.} originates from the existence of the on-shell Lorentz covariant and supersymmetric candidate counterterms, see for example \cite{Kallosh:1980fi, Bossard:2011tq, Freedman:2017zgq}.
They are often presented as some integrals over the superspace, or sub-superspace based on Lorentz covariant superfields associated with  supermultiplets shown in Table \ref{Table_1}, Table \ref{Table_2}, Table \ref{Table_3} for $\cN=8,6,5$ respectively.

Let us look, for example, at the 3-loop supersymmetric linearized $R^4$ counterterm in the form given in  \cite{Freedman:2011uc}. It has  51 different terms for the 4-point candidate UV divergence. One of them is a  2-vector-2-graviton 4-point candidate for an UV divergence
\be \label{CT}
\mathcal{L}_{CT}=C^{\alpha\beta\gamma\delta} \bar C^{\dot{\alpha}\dot{\beta}
\dot{\gamma}\dot{\delta}}  \nabla_{\alpha\dot{\delta}}
 F_{\beta\gamma AB} \nabla_{\delta\dot{\alpha}} \bar {F}_{\dot{\beta}\dot{\gamma}}^{ AB} \; .
\ee
It depends on $\bf 28$ and $\overline{\bf 28}$ of $SU(8)$ vector field strengths and their conjugates and $\bf 70$ scalars. The counterterm  is manifestly supersymmetric at the linearized level \cite{Kallosh:1980fi}, all 51 terms are packaged in a superspace expression  $\int d^4 x d^{16} \theta [W]^4$ in a particular basis in superspace, where the superfield $W(x, \theta)$ represents the  graviton supermultiplet in Table \ref{Table_1}. If the relevant UV infinity were to occur, it would mean that \E\ symmetry current conservation must be broken  \cite{Kallosh:2011dp}.
The BN proposal is to promote the counterterm to the status of the manifestly \E\, invariant SoD. For this purpose one should  promote  each
of the $\bf 28$ and $\overline{\bf 28}$ of $SU(8)$ to a graviphoton $T_{  \alpha\beta\, AB}$ and $\bar{T}_{\dot\alpha\dot\beta}^{ AB}$,
which makes the expression in \rf{CT} a functional of the
double set of $\bf 28$ and $\overline{\bf 28}$ of $SU(8)$ :
\bea\label{SoD1}
\mathcal{L}_{CT} \qquad &\Rightarrow & \qquad  {\rm SoD} = C^{\alpha\beta\gamma\delta} \bar C^{\dot{\alpha}\dot{\beta}
\dot{\gamma}\dot{\delta}}  \nabla_{\alpha\dot{\delta}}
 T_{\beta\gamma AB} \nabla_{\delta\dot{\alpha}} \bar {T}_{\dot{\beta}\dot{\gamma}}^{ AB} \, , \\
 \bf 28 \quad \& \quad \overline{\bf 28} \qquad &\Rightarrow&  \qquad  \qquad 2\times ( \bf 28 \quad \& \quad \overline{\bf 28})\, .
 \eea
 Note that  the graviphoton
\be
 T_{\beta\gamma AB} =  h_{\Lambda
 AB}\, F^{\Lambda \beta\gamma} - f^\Lambda_{AB} \,G_{ \beta\gamma\Lambda} \label{gp} \ee
  in SoD depends on both $F^{\Lambda \beta\gamma}$ and $G_{ \beta\gamma\Lambda}$ which are independent of each other, as required by E7 invariance, this is why we have to deal with the doubling of vector fields. This is different from the expression in \rf{CT} for the candidate UV divergence, where  $G^+ = \cN F^+$ and $G^-= \overline {\cN} F^-$ and it is $SU(8)$ invariant and  there is a supersymmetric extension of the expression in \rf{CT}, as given explicitly at the linear level in \cite{Freedman:2011uc}.

Moreover,   note  that in the BN proposal  the symplectic section, $(h_{\Lambda AB},  f^\Lambda_{AB})$ which is used to build the SoD according to eqs.  \rf{SoD1}, \rf{gp}, does not change from the classical case, scalars must remain the same to preserve the classical coset structure.

To find out if the supersymmetric extension of the bosonic SoD is available for  all $\cN\geq 5$ theories, we have studied  all supermultiplets with a double set of {\it dynamical} vector fields \footnote{We  have explained in Sec. 3 why, based on all known constructions of supersymmetric theories,  it is not possible to realize the BN deformation proposal \cite{Bossard:2011ij}, using auxiliary vector fields within the standard framework of extended supergravity theories.}
as we need in the bosonic SoD in eq. \rf{SoD}. The results follow simply  from the unitary representations of  the conformal Lie superalgebra $SU(2,2|\cN)$ . We presented all possible supermultiplets which could be used in supersymmetrization of the SoD given in \rf{SoD}. We proved that it is not possible to double the vectors without changing the basic underlying premises of the theory such as the  number and representations of $SU(\cN)$ for the scalars, and to avoid higher spins, in all available supermultiplets of the corresponding superconformal algebra.

The authors of \cite{Bossard:2010dq} studied the perturbative quantization of  $\cN=8$  supergravity in a formulation in which the full U-duality group $E_{7(7)}$ is realized off-shell and which is not manifestly Lorentz invariant. They showed that  $E_{7(7)}$ anomalies cancel as a consequence of the vanishing $SU(8)$ anomalies \cite{Marcus:1985yy} to all orders in perturbation theory. We expect these results to extend to $\cN=6$ and $\cN=5$ supergravities as well. 
Therefore our results suggest that simple E7 type duality symmetry in $\cN\geq 5$ supergravity theories together with   supersymmetry might protect $\cN \geq 5$ supergravity from UV divergences assuming supersymmetry  does not become anomalous at higher loops and the results of \cite{Bossard:2010dq} hold for manifestly  Lorentz invariant formulations.
This reasoning  is supported by  the established  UV finiteness of  $\cN=8$ and $\cN=5$ at 3 and 4 loops. Particularly important here is the case of $\cN=5$ at 4 loops. Until the recent paper  \cite{Kallosh:2018wzz}, not a single explanation of $\cN=5$ UV finiteness in  4 loops  was proposed. Here, we find
that the situation with candidate counterterms which break duality \cite{Kallosh:2011dp} is as follows. The proposal suggested in  \cite{Bossard:2011ij} to restore duality was based  on the assumption that also supersymmetry is unbroken in the deformation process. Our study of this issue led us to the conclusion that BN deformation procedure can not  be consistent with supersymmetry within the standard framework of supergravity theories. From this perspective the UV finiteness of  $\cN=8$ and $\cN=5$ at 3 and 4 loops is  a consequence of the fact
that that both supersymmetry and duality are not anomalous. In such case, together, they predict UV finiteness of $\cN\geq 5$ supergravity, as we argue here on the basis of the absence of a consistent deformation preserving duality and supersymmetry in presence of UV divergences. Thus our analysis, based on all known constructions in supergravity,  suggests the following: unbroken duality and supersymmetry forbid UV divergences, and in case of $\cN=8$ and $\cN=5$ at 3 and 4 loops the
 computations in \cite{Bern:2007hh,Bern:2014sna} support our interpretation that both of these symmetries are respected in these perturbative computations.

The argument about absence of SoD which we gave here, based on unbroken duality and supersymmetry and within the standard assumptions about extended supergravity, is valid at any loop order in  $\cN\geq 5$ supergravity. Whether these symmetries continue to be respected in higher loops, i. e. whether $\cN\geq 5$ supergravity theories remain fully duality invariant, supersymmetric and anomaly-free,  remains to be seen.

\section*{Acknowledgement}
We are grateful to  B. de Wit, S. Ferrara, D. Freedman, A. Linde,  H. Nicolai,  R. Roiban, A. Tseytlin  and Y. Yamada for the important  discussions. RK is supported by Stanford Institute for Theoretical Physics and by the US National Science Foundation Grant  PHY-1720397.
MG  would like to thank the hospitality of  SITP where  this work was performed.

\appendix

\section{ $\cN=2$ supergravities interacting with matter}
Above we have  studied $\cN\geq 5$ supergravity models, which have no matter  multiplets, only  gravitational ones. In  Sec.  3 we considered a truncation of $\cN= 8$ supergravity to $\cN=2$ Maxwell-Einstein supergravity theory describing the coupling of 15 vector multiplets,  the quaternionic magical supergravity theory \cite{Gunaydin:1983rk,Gunaydin:1983bi}. This was done with the purpose to study the issue of auxiliary fields in $\cN= 8$ by looking at its  truncated version.

Here we would like also to add few more comments about the general case of $\cN=2$ supergravity coupled to vector multiplets.
Matter-coupled supergravities are expected to have one-loop UV divergences depending on the matter energy-momentum tensor $R_{\mu\nu} - {1\over 2} g_{\mu\nu} R=T_{\mu\nu}^{ \rm mat}$. The relevant one-loop UV divergence is
 $(T_{\mu\nu}^{ \rm mat})^2 +\cdots $. In pure supergravities $\cN\geq 5$ this one-loop gauge-invariant UV divergence is absent since $R_{\mu\nu} - {1\over 2} g_{\mu\nu} R=0$ on shell for all $\cN\geq 5$ supergravities.

The magical supergravity theories  were discovered long time ago \cite{Gunaydin:1983rk,Gunaydin:1983bi}. They are defined by the four simple Jordan algebras of degree three realized by $3 \times 3$ Hermitian matrices over the four division algebras.  Their global symmetry groups in five, four and three dimensions correspond to the groups that appear in the Magic Square of Freudenthal, Rozenfeld and Tits. Hence the name.
The $d=4$ quaternionic magical supergravity can be truncated to the complex and real magical $\cN=2$ supergravity theories with 10 and 7 vector fields. The octonionic magical $\cN=2$ supergravity theory with 28 vector fields can not be embedded in $\cN=8$ supergravity.
The U-duality groups of the four magical supergravity theories in $4d$ are all groups of type E7, namely  $E_{7(-25)} , SO^*(12) , SU(3,3)$ and $Sp(6,\mathbb{R})$ under which the vector field strengths and their magnetic duals transform in 56, 32, 20 and 14 dimensional representations, respectively. Among all the $\cN=2$ Maxwell-Einstein supergravity theories with homogeneous scalar manifolds they are distinguished by the fact that their U-duality groups are simple and the vector field strengths and their magnetic duals  form a single irreducible symplectic representation. This is a property  they share with $\cN\geq 5$ supergravity theories.

However unlike $\cN \geq 5$  supergravities generic N=2 Maxwell-Einstein supergravity theories with homogeneous scalar manifolds have one loop divergences \cite{Carrasco:2012ca,Chiodaroli:2015wal,Chiodaroli:2017ngp,Ben-Shahar:2018uie}. As was pointed out in \cite{Ben-Shahar:2018uie} these divergences correspond to two independent linearized  counterterms and the divergences associated with one of these counterterms are absent {\it only } for the magical supergravity theories \cite{Gunaydin:1983rk,Gunaydin:1983bi}, which have simple U-duality symmetry groups of type E7. The first UV divergence in \cite{Ben-Shahar:2018uie} corresponds to the term $(T_{\mu\nu}^{ \rm mat})^2 +\cdots $ which we discussed above. It is duality invariant since the energy momentum tensor is duality invariant. The fact that in magical supergravities the second type of UV divergence vanishes might be a consequence of E7 type duality, but this requires a separate investigation, especially if it persists at higher loops.

 $\cN=2$ theories  have axial
 and conformal anomalies which are absent for $\cN \geq 5$ supergravities \cite{Marcus:1985yy, Meissner:2016onk, Kallosh:2016xnm} whose one-loop  amplitudes are also anomaly-free \cite{Freedman:2017zgq}. This  implies that the U-duality groups of type E7 of the magical $\cN=2$ supergravity theories might be broken at the quantum level. Hence the argument that the finiteness of $N=6$ supergravity may be understood  as a consequence of exact $SO^*(12)$ U-duality symmetry at a given loop order can not be extended to the magical quaternionic $\cN=2$ supergravity  with the same bosonic content since the anomalies tend to break  U-duality symmetry at the quantum level already at one loop level. This is suggested by the studies of $\cN=4$ supergravity with and without matter at the one-loop level where  there are anomalies, as shown  in \cite{Carrasco:2013ypa}, and at the four-loop level, where there are related UV divergences, as shown in
\cite{Bern:2013uka}. More recent developments in $\cN=4$ supergravity  in   \cite{Butter:2016mtk}  and in  \cite{Bern:2017rjw} suggest that the relation between anomalies and UV divergences in extended supergravities might be more interesting, and new insights can be expected.

\providecommand{\href}[2]{#2}\begingroup\raggedright\endgroup



\end{document}